\documentclass[sn-mathphys,Numbered,iicol]{sn-jnl}%

\usepackage{graphicx}%
\usepackage{amsmath,amssymb,amsfonts}%
\usepackage[title]{appendix}%
\usepackage{xcolor}%
\usepackage{manyfoot}%
\usepackage[inline,shortlabels]{enumitem}

\usepackage[compact]{titlesec}
\titlespacing{\section}{0pt}{2ex}{1ex}

\titlespacing{\subsection}{0pt}{1ex}{0ex}
\titlespacing{\subsubsection}{0pt}{0.5ex}{0ex}

\begin{document}

\title[Generative Optimization]{Generative Optimization: A Gateway to AI-Enhanced Problem-Solving in Engineering}

\author[1]{\fnm{Lyle} \sur{Regenwetter}}
\author*[1]{\fnm{Cyril} \sur{Picard}}\email{cyrilp@mit.edu}

\author[1]{\fnm{Amin} \sur{Heyrani Nobari}}%

\author[2]{\fnm{Akash} \sur{Srivastava}}%

\author[1]{\fnm{Faez} \sur{Ahmed}}%

\affil[1]{\orgdiv{Massachusetts Institute of Technology}, \orgaddress{\city{Cambridge}, \postcode{02139}, \state{Massachusetts}, \country{USA}}}

\affil[2]{\orgdiv{IBM}, \orgaddress{\city{Boston}, \postcode{02210}, \state{Massachusetts}, \country{USA}}}

\abstract{
The field of engineering is shaped by the tools and methods used to solve problems. Optimization is one such class of powerful, robust, and effective engineering tools proven over decades of use. Within just a few years, generative artificial intelligence (GenAI) has risen as another promising tool for general-purpose problem-solving. While optimization shines at precisely identifying highly-optimal solutions, GenAI excels at inferring problem requirements, bridging solution domains, handling mixed data modalities, and rapidly generating copious numbers of solutions. These differing attributes also make the two frameworks complementary. Hybrid `generative optimization' algorithms have gained traction across a few engineering applications and now comprise an emerging paradigm for engineering problem-solving. We expect significant developments in the near future around generative optimization, leading to changes in how engineers solve problems using computational tools. We offer our perspective on existing methods, areas of promise, and key research questions. 
}

\keywords{optimization, generative AI, generative optimization, engineering design}

\maketitle

\section{Introduction}\label{sec1}

Whether searching for stronger metal alloys, safer pharmaceutical treatments, or more efficient wind turbines, engineers seek ever-improving solutions to pressing real-world challenges. To support this advancement, the engineering process itself is also ever-improving, thanks to continual progress in tools and methods~\cite{cagan_framework_2005}. Computational approaches like numerical simulations have replaced physical prototypes and expert intuition to quickly evaluate more possible solutions~\cite{ulrich_product_2020}. However, with faster evaluation, the search for new possibilities to evaluate becomes a challenge in its own right. For decades, optimization has been the classic technique for automating this process, thereby solving countless problems in various engineering disciplines and beyond~\cite{coello_coello_evolutionary_2007}. However, to tackle more complex systems with increasingly strict requirements, engineers must continue to revisit and upgrade their search toolboxes.

Optimization refers to the process of systematically adjusting a fixed set of solution variables to maximize some explicit objectives, subject to explicit constraints. 
Although standard, the rigidity of the solution representation, objectives, and constraints lead to limitations. Consider the optimization of a wind turbine blade to maximize aerodynamic efficiency subject to a variety of manufacturing constraints.
Rigidly representing such a complicated problem in order to use optimization can require significant domain expertise~\cite{mdobook} or simplifying assumptions~\cite{ryerkerk2017solving}. Can a fixed number of variables define the smooth continuity of a turbine blade? How would a static objective capture aerodynamic efficiency when weather and operating conditions continuously vary? What limited set of constraints could possibly regulate the limitless combinations of possible manufacturing processes? The need for rigid structure limits optimization's adoption in industry and its adaptation to new problem variants. Optimization methods can also struggle to generate high-quality and diverse solutions~\cite{weise2009optimization}, including when the solution space is large~\cite{zhou2024evolutionary, binois2022survey}, or highly constrained~\cite{picard_realistic_2021}, conditions common in systems engineering. Though many of these shortcomings can be tackled with specialized optimization approaches, they also present an opportunity for entirely different classes of computational tools to shine. 

Generative Artificial Intelligence (GenAI) is another class of computational tool increasingly used to search for solutions in engineering~\cite{regenwetter2022deep, shin2023topology, cheng2021molecular, anstine2023generative, liao2024generative}.
GenAI has been popularized in recent years by breakthroughs in text and image synthesis~\cite{kingma2014, goodfellow2014generative, ho2020denoising} (e.g., GPT-4 for language modeling, DALL·E and Stable Diffusion for image generation). Given the complex representations and massive solution spaces common in these domains, many wonder whether GenAI is the future of generalizable problem solvers~\cite{fui2023generative}.  

GenAI formally refers to a class of machine learning models that learn a probability distribution over a training dataset and generate new samples from this learned distribution, which then generally resemble the training data. Critically, GenAI models rely on representation-learning mechanisms to autonomously create an internal representation of solutions, often called an embedding. Popular GenAI methods include variational autoencoders, generative adversarial networks, transformer encoders-decoders, and diffusion models.

Contrary to numerical simulations, surrogate models, or digital twins that have dramatically improved how engineers analyze and model systems, GenAI is primarily used as a tool to search for and propose possible solutions.
Indeed, in engineering contexts, GenAI models have been trained on existing solutions (e.g. known drug molecules,  mechanical structures, product designs, etc.) to capture underlying statistical patterns. Once trained, they can then instantaneously propose new molecules~\cite{cheng2021molecular, bilodeau2022generative}, structures~\cite{nie2021topologygan, maze2023diffusion}, and products~\cite{oh2019deep, regenwetter2022framed} as potentially novel solutions. Specifically, GenAI addresses several of optimization's key limitations through:

\begin{itemize}
    \item \textbf{Adaptability:} Learning variable solution representations~\cite{achiam2023gpt,Li2022Deep, Song2023Multimodal} and implicit requirements~\cite{regenwetter2025multi},  allowing for more flexible application to complex problems.
    \item \textbf{Speed:} Generating hundreds of candidate solutions in seconds once trained~\cite{nobari2024link, watson2023novo}, enabling rapid exploration of large solution spaces.
    \item \textbf{Diversity:} Producing varied solutions through probabilistic sampling of the learned solution space, sometimes uncovering unconventional solutions~\cite{chen2021padgan, regenwetter2023beyond}.
\end{itemize}

Nonetheless, challenges abound as well~\cite{regenwetter2023beyond}. GenAI's probabilistic nature means it may violate hard constraints~\cite{regenwetter2024constraining} (e.g., manufacturability rules), struggle to adapt to cases outside or at the edge of its training data~\cite{woldseth2022use}, and require large datasets that are often costly to acquire in engineering domains~\cite{10.1115/DETC2023-111609,10.1115/1.4067871}.

In many ways, optimization and GenAI are markedly dissimilar: optimization identifies exact solutions to precisely-defined problems, while GenAI randomly samples solution candidates from abstract probabilistic representations. Yet, there is an opportunity for synergy through \textbf{\textit{generative optimization}}, a concept we define as any algorithmic framework that couples components of GenAI with components of explicit optimization. 
Within generative optimization, for example, GenAI can explore and represent possible solutions and translate multimodal requirements, while optimization can evaluate these possibilities against objectives and constraints, train and constrain GenAI, or perform subselection---yielding fast, requirement‑satisfying engineering solutions.
Exploration into generative optimization is seeing rapid growth, mirroring the increasing traction of GenAI in engineering domains. In this perspective, we articulate what generative optimization entails, highlight existing efforts to harness it, and sketch out its role in engineering applications from promising preliminary results to underexplored areas.

\begin{figure*}
    \centering
    \includegraphics[width=\textwidth]{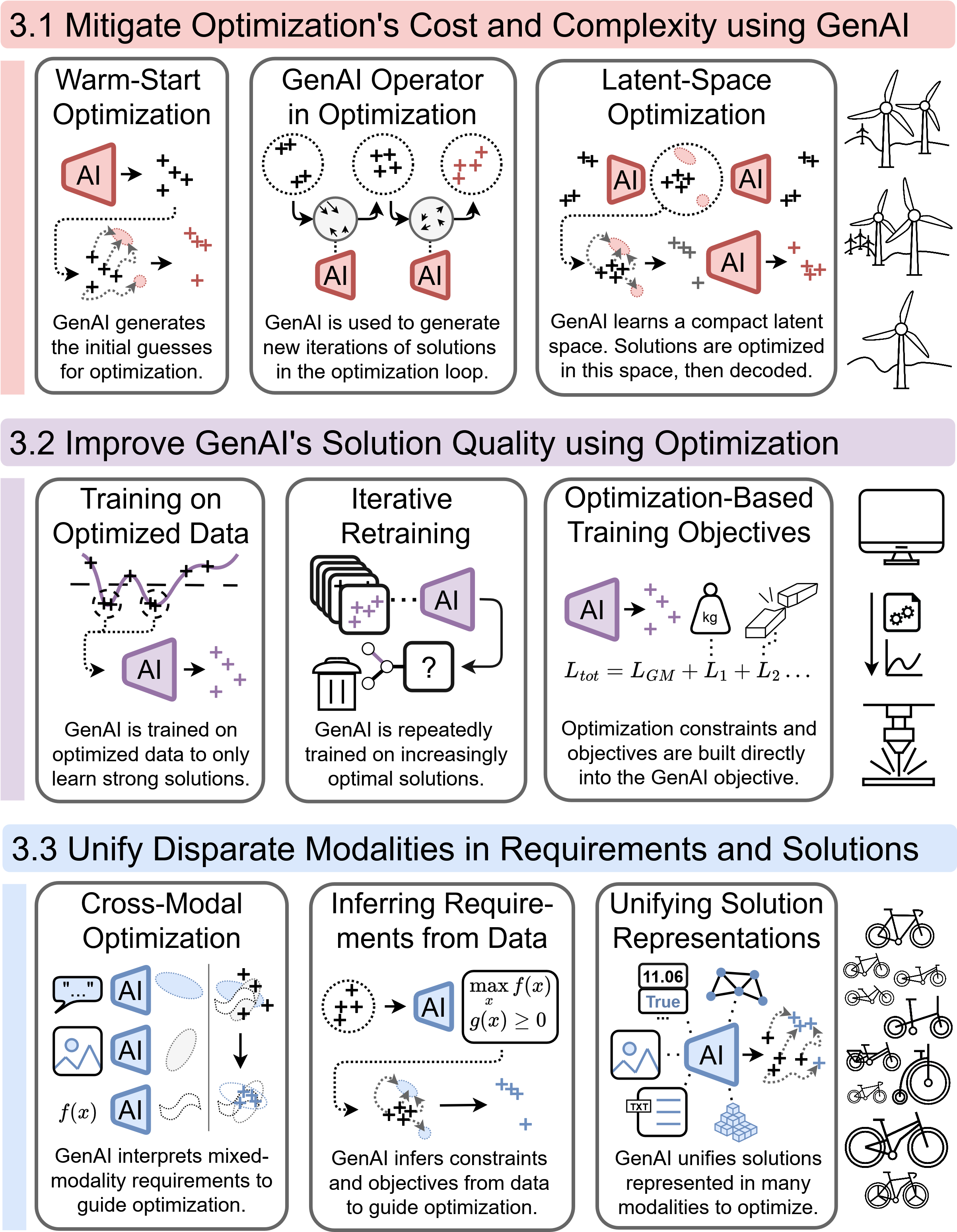}
    \caption{Overview figure showing different techniques to combine Generative AI and Optimization, along with pictograms of the example cases discussed in each section. These approaches are grouped into methods that seek to mitigate optimization's cost and complexity using GenAI, improve GenAI's solution quality using optimization, or unify disparate modalities in requirements and solutions. }
    \label{fig:overview}
\end{figure*}

\section{Discussion}

Current research on integrating GenAI with optimization is organized into three primary categories:
\begin{enumerate*}[1)]
    \item Using GenAI to expand the speed and generalizability of optimization;
    \item Using optimization to enhance the solution quality of GenAI;
    \item Learning or inferring objectives, constraints, or solution representations for optimization using GenAI.
\end{enumerate*}
For each approach, we present a motivating example, briefly highlight existing strategies, and finally suggest new avenues that we believe have significant potential. Fig.~\ref{fig:overview} provides an overview of the structure of this discussion.

\subsection{Mitigate Optimization's Cost and Complexity using GenAI}
During most optimization processes, many potential solutions are evaluated. Since their evaluation may involve costly physical experimentation or computational simulations, advances in optimization algorithms have generally sought to identify better solutions in a more sample-efficient manner. As we discuss in this section, generative AI has the potential to enhance and accelerate optimization in domains where prior data is available or can be collected.

In this section, we consider a motivating example of a hypothetical company that specializes in wind turbine layout optimization. Due to the complex physical phenomena like aerodynamic wake interactions~\cite{chen2016wind, feng2017design} geographic topography~\cite{zilong2022layout} and weather~\cite{chowdhury2013optimizing}, this problem is extremely costly to model and optimize~\cite{elkinton2008algorithms, rodrigues2016multi}. This company faces two conflicting challenges: Modeling costs are extreme, so they can only afford to evaluate a few turbine layouts for a given contract. Simultaneously, the solution space is large and complex, requiring many evaluations to converge on a high-quality solution. Therefore, the company seeks to leverage GenAI to reuse insights from past wind farm projects, improving the cost and complexity of their optimization procedure.
This section discusses several avenues to leverage GenAI to either propose strong candidate solutions or simplify the solution space, making the optimization process more manageable. Since any upfront training and data collection cost is ideally repaid by faster future optimizations, this approach is sometimes referred to as amortized optimization~\cite{amos2023tutorial}.

\subsubsection{Warm-Start Optimization}
The efficiency and success of optimization often depend on the initial guess. Compared to the random solutions that are often used for initialization, AI-generated solutions may be closer to the optimal solution, requiring fewer optimization iterations to converge.
For example, the wind farm company might generate better initial guesses using a GenAI model trained on data from previous turbine layout projects. Compared to plain optimization, this ``warm-starting'' is often able to reach stronger ultimate solutions, while reaching comparable solutions in as few as 10\% as many optimization steps~\cite{chen2022inverse}. Its effectivness has been demonstrated in the generation of mechanical structures~\cite{maze2023diffusion, JooYuJang2021}, airfoil design~\cite{chen2022inverse}, and motion planning~\cite{huang2024diffusionseeder}.

While warm-starting can be performed using non-generative AI models~\cite{baker2019learning, chen2022large, nobari2024nito}, generative models have several advantages in inverse design settings, such as improved generalization. 
Nonetheless, GenAI models can still introduce biases based on their training data. This can potentially cause the optimization to miss better solutions that lie outside the model's learned distribution. Due to the complex interplay between optimization success and initial guesses~\cite{10.1115/1.4063006}, 
certain properties of the GenAI model, training data, or optimization algorithm can be engineered to improve warm starting~\cite{nobari2024nito}, and this presents an open area of investigation.

\subsubsection{GenAI Operator in Optimization}\label{ml-assisted}
Instead of giving optimization algorithms a head-start, GenAI can also be incorporated directly into the optimization algorithm to guide its search process~\cite{saxena_machine_2024}. This differs from surrogate-assisted optimization where the optimization algorithms remain unchanged while the objective or constraint functions are replaced by a surrogate. In GenAI-assisted optimization algorithms, the optimization procedure itself is changed to leverage GenAI to accelerate convergence~\cite{mittal_learning-based_2021} or promote diversity~\cite{saxena_machine_2024}. 

One such approach may augment the classic mutation and crossover operators used in many evolutionary optimization algorithms with a data-driven model, which suggests solutions that are likely to be more optimal~\cite{mittal_learning-based_2021, shem2025deep}. Alternatively, augmenting such operators with a distributional density estimator, the backbone of GenAI, has been shown to significantly enhance constraint satisfaction in optimization~\cite{regenwetter2025multi}. Using these types of approaches, the wind farm contractor can focus resources on designs that are more promising based on reasoning or trends from existing data, rather than spending time and money evaluating randomly-generated designs that are many times~\cite{regenwetter2025multi} more likely to violate constraints. 

To make strong recommendations, however, bespoke GenAI-based operators may have to be updated during the optimization as more information is gathered~\cite{surina2025algorithm}. This can be challenging due to training costs and the risk of data bias. Instead of bespoke models, pretrained foundation models with generalizable knowledge can also be leveraged as operators. For example, large language models have demonstrated strong enough reasoning capabilities to act as `smart' optimization operators~\cite{liu2024large}, an approach which has seen applications in engineering design~\cite{zhang2025using} and even achieved state of the art performance in a variety of combinatorial optimization problems~\cite{romera2024mathematical, grayeli2024symbolic}, such as the famous Traveling Salesman Problem~\cite{ye2024reevo}. However, to avoid the overhead of translating optimization solutions to and from language, pretrained parametric foundation models such as prior-data fitted networks~\cite{muller_pfns4bo_2023} could conceivably be directly deployed as optimization operators by continually ingesting new samples and suggesting new solutions during the optimization process. Regardless, harnessing GenAI's distributional learning capabilities to recommend highly promising solutions can allow optimization to identify more optimal solutions~\cite{romera2024mathematical} using just a fraction of the computational cost~\cite{shem2025deep}. 

\subsubsection{Latent-Space Optimization} 
Optimizing just a few variables at once is typically easier than simultaneously optimizing hundreds or thousands~\cite{wes-4-663-2019}. Due to hidden relationships between variables, high-dimensional optimization problems can often be simplified to make optimization more efficient. Although this simplification can be performed using domain expertise~\cite{wes-4-663-2019}, many GenAI models can instead examine data to capture these hidden relationships. In doing so, the GenAI model learns a compact set of `latent' variables which it can map to the original real-world variables. Then, instead of directly optimizing the original variables, the optimizer can more efficiently explore and adjust latent variables due to their lower dimensionality and compactness~\cite{grantham_deep_2022,wang2022phononic}. The GenAI model is then able to map any latent variable solutions to the original high-dimensional solution space.

In practical applications, for instance, optimizing the layout of wind turbines might involve hundreds of variables such as position, height, and angle for numerous turbines. Since optimizing fine-grained details of each turbine may be challenging and computationally expensive, the company can instead use a GenAI model to learn a latent representation of turbine layouts based on historical data. This learned latent space encapsulates essential aspects of the layouts with fewer dimensions and provides a more manageable framework for optimization.
This approach has been successfully applied in fields like molecular design~\cite{kusner2017grammar, jin2018junction, gomez2018automatic, liu2018constrained}, structural optimization~\cite{guo2018indirect}, material design~\cite{wang2022phononic, yang2018microstructural, xue2020machine} and other engineering problems~\cite{zhang20193d, chen2019aerodynamic, rios2021multitask}, where complex, multidimensional problems benefit significantly from such dimensional reduction and efficient exploration strategies.

While latent optimization offers significant advantages in simplifying complex optimization tasks, its effectiveness is contingent on the compactness~\cite{eissman2018bayesian, wang2020deep}, smoothness~\cite{lee2023advancing, chen2019aerodynamic}, disentanglement~\cite{lew2021encoding, Fujita2021Design}, representational capacity~\cite{tripp2020sampleefficient, gomez2018automatic}, and robustness~\cite{tripp2020sampleefficient, jin2018junction} of the latent space that the underlying GenAI model encodes. Ongoing research continues to address the many areas of GenAI latent space quality through better GenAI models, or by developing new latent optimization frameworks to better leverage latent spaces for optimization~\cite{chu2024inversionbased, moss2025return}.

\subsection{Improve GenAI's Solution Quality using Optimization}
Having explored methods for GenAI to accelerate and simplify optimization methods, we now examine methods to completely replace optimization with a GenAI model whose training is informed by the optimization process it replaces. This optimization-informed training addresses one of GenAI's key shortcomings in engineering -- its classic inability to generate requirement-satisfying solutions~\cite{regenwetter2022deep, regenwetter2023beyond}. Optimization-assisted GenAI is particularly valuable in domains requiring solutions in real time or solutions to millions of unique problems where cumulative computational costs would be prohibitive.

Digital manufacturing is an application area where such fast or iterative problem solving tasks are common. For example, in dynamic manufacturing, process parameters are precisely customized to optimize part quality~\cite{shen2019learning, mattera2024optimal, kang2009virtual}, or even adjusted on the fly in response to monitoring data~\cite{zhang2019process, park2020digital, gunasegaram2024machine}, requiring instantaneous decision making. Or consider computer-generated machine toolpaths, which must be re-optimized in an error-prone and expensive iterative process~\cite{10.1016/j.engappai.2023.106464} for every unique part. GenAI can support new digital manufacturing technologies where optimization would result in unacceptable latency or amortized cumulative costs. GenAI has already seen successful use in modeling uncertainty to predict the space of plausible manufacturing outcomes, given a particular set of process conditions~\cite{kim2023virtual, mu2024online}. Extending the GenAI model to directly supply optimal process conditions or control plans is a natural extension of such methods. 

In this section, we consider a hypothetical Digital Manufacturing (DM) company seeking to improve the software of their machines. Given a part to manufacture, this software selects optimal configurations of process parameters (temperature, laser power, speed, toolpath, etc.) to minimize residual stress and other manufacturing outcomes using a conventional optimization process. The company wants to replace this optimization with a GenAI model, which would not only eliminate several minutes of optimization per part, but allow their machines to dynamically update process parameters on the fly if the machine detects issues. To ensure that the AI model generates process parameters that yield high-quality parts, they seek to transfer knowledge from their optimizer into the GenAI model during training. 

\subsubsection{Training GenAI on Optimized Data}
GenAI models tend to outperform optimization in solution speed, but lag behind in solution quality~\cite{woldseth2022use, regenwetter2025bike}. Realizing this, the company is happy to invest significant effort into improving the GenAI model's performance, even if this means investing time in training and data-generation up-front. A reliable way to improve a GenAI model's performance is to improve the quality of its training data, which can be done using optimization.  

A representative example, which has been extensively explored, is GenAI-based generation of optimal free-form structures~\cite{rawat2019application,sharpe2019topology,guo2018indirect, nie2021topologygan, maze2023diffusion, 10.1115/1.4062980}, in which a well-established optimizer~\cite{bendsoe1989optimal} is used to generate the GenAI training dataset. Analyses of such approaches generally compare them to a baseline of pure optimization~\cite{woldseth2022use, shin2023topology}. A common concern is generalizability, where the optimized training data may lack the diversity to result in a robust GenAI model~\cite{woldseth2022use}. Another  is the optimization cost of GenAI training data, which may be difficult to amortize~\cite{woldseth2022use}. However, when inference-time cost is drastically more important than up-front cost, such as in the DM example at hand, amortization is less of a concern. In addition to better inference-time cost, GenAI trained on optimized solutions has notable capabilities in solution space exploration and diverse generation~\cite{shin2023topology, regenwetter2023beyond}. Training GenAI on optimized data has been explored in a variety of other engineering domains, including photonics~\cite{jiang2019free, liu2021tackling}, aerodynamics~\cite{chen2022inverse}, and more~\cite{gaier2024generative}. 

Partially-optimized solutions can also be leveraged by sequential GenAI models, like diffusion or autoregressive models. For example, these models can align their intermediate outputs with the steps taken by gradient-based~\cite{giannone2023aligning} or evolutionary~\cite{yan2024emodm} optimization. Using such techniques, the DM company can teach their generative model to iteratively improve process parameters or machining toolpaths by following an optimization trajectory. After supervising the training using the expensive simulations, the model would ideally learn to mimic optimization trajectories independently, eliminating this inference-time cost. 

\subsubsection{Iterative Retraining}
The concept of training GenAI models on optimized solutions can be extended to progressively train successive models on better and better solution sets. For example, the DM company may train a model, generate many process configurations, evaluate them, and then select a subset of only the most optimal generated configurations to train the next iteration of their GenAI model. By repeating this process, the company can theoretically train better and better models that generate progressively stronger process parameter solutions. Such ``iterative retraining'' approaches have been explored across a variety of engineering applications~\cite{shu20203d, oh2019deep, jiang2019free, wang2022phononic} and are often performed in conjunction with latent-space optimization~\cite{abeer2024multi, tripp2020sampleefficient}.  

Absent proper precautions, however, iterative retraining may result in ``mode collapse,'' meaning that diversity is quickly lost in successive iterations. To combat this, re-optimization of generated solutions at each iteration can help improve diversity in successive model instances~\cite{jiang2019free}, an approach that can just as accurately be described as optimization with GenAI in the loop (as in Sec.~\ref{ml-assisted}). The cost to repeatedly retrain the GenAI model is another significant concern.  This can conceivably be addressed using a retraining-inspired, but nonetheless training-free approach using in-context-learning, where the context of an foundation model is iteratively updated with progressively better solution sets. Although minimally explored, this approach is a natural extension of retraining to the increasingly popular in-context-learning paradigm~\cite{dong2022survey}.  

\subsubsection{Optimization-Based Training Objectives}
GenAI models are typically trained to optimize for statistical similarity to a training dataset, irrespective of the quality of solutions in the dataset~\cite{regenwetter2023beyond}. This statistical similarity objective can be augmented with a optimization-based performance-indicating function~\cite{dong2020inverse, chen2021padgan, chen2021mopadgan, pcdgan, regenwetter2022design, maze2023diffusion} to encourage the generation of higher-performing solutions. The hypothetical DM company could even borrow the heuristics used by their current optimizer to evaluate the quality of process configurations. This could push their GenAI model to synthesize process parameters of competitive quality to their optimizer.

Differentiability is a key requirement of any training objective used in GenAI~\cite{chen2021padgan}. If the optimization-based training objectives are non-differentiable, a differentiable data-driven estimator can be trained instead. However, trained evaluators may struggle with generalizability and exact evaluators may be computationally expensive~\cite{bagazinski2024c}, presenting ongoing challenges with this approach. Alternatively, performance-based conditioning can be used to encourage higher-performing solutions~\cite{pcdgan, nobari2022range}. Overall, the incorporation of optimization-based training objectives has significantly improved both the quality and validity of GenAI-generated solutions in a variety of engineering domains, including materials design~\cite{dong2020inverse}, structural mechanics~\cite{maze2023diffusion}, and more~\cite{regenwetter2022design}. 

\subsection{Unify Disparate Modalities in Requirements and Solutions}
As a third broad topic, we address one of the key usability limitations of optimization in practice~\cite{saadi2023generative, fajemisin2024optimization, saadi2024effect}: Its dependence on a rigidly-defined solution space, set of objectives, and set of constraints. Generative modeling and representation learning unlock new possibilities to mitigate this dependence and apply optimization to new problems. Among the widespread ramifications, the impact on  democratization and customization, for example, is easily appreciated. Consider bicycle design, a domain with a variety of technical design objectives~\cite{wilson2020bicycling}, compatibility requirements with a rider's morphology~\cite{burt2022bike}, and an intrinsic dependence on human preference~\cite{hoor2022bicycle}. 
Designs would ideally be customized not only to accommodate individual users' objective needs and match their morphology, but also to reflect their individual aesthetic preferences. Although mass customization has historically led to competitive advantages in the bicycle industry~\cite{kotha1996mass}, it can be limited by design throughput. The hundreds of design attributes~\cite{regenwetter2021biked} and dozens of requirements~\cite{regenwetter2025bike} make the design of each bicycle a significant endeavor. 

As a motivating example for this section, we consider a bicycle manufacturer that wants to release a customization and optimization tool that would give individual cyclists maximum flexibility to customize their bikes, while automating the technical design, performance optimization, and engineering constraint validation. To do so, the bike manufacturer wants to empower users to specify their personal design requirements intuitively, perhaps through a text description of their needs, a photo of themselves, or an model of an existing design for inspiration. Their AI-based system would process users' vague or non-technical requests into precise, complete, and actionable design requirements, while inferring any constraints that their laymen customers may forget to explicitly specify. Finally, they need a system
to translate between multimodal solution representations during optimization to arrive at a final optimal design, described using a representation compatible with their manufacturing processes.

\subsubsection{Cross-Modal Optimization}
Current direct optimization approaches are generally only trusted to handle easily quantifiable requirements~\cite{saadi2024effect}. Generated solutions then need to be manually modified to accommodate unquantified requirements, which can require significant expertise~\cite{saadi2023generative} and time~\cite{saadi2024effect}.
To overcome this shortcoming, GenAI models may enable humans to interact with optimization algorithms more flexibly and intuitively. LLMs, for example, are becoming increasingly capable of processing natural language into 
formal logic~\cite{cosler2023nl2spec, li2025extracting}, engineering constraints~\cite{yan2025assertllm, he2024utilizing}, and
system requirements~\cite{li2025specllm, doris2025designqa}. 
These capabilities can be applied to interpret textual constraints into the rigid mathematical objectives and constraints used in optimization~\cite{qin2024intelligent, austin2024bayesian, zhang2025autolead}, though significant gaps remain~\cite{abdollahi2024hardware}. 

Foundation models for cross-modal representation learning can also enable optimization using highly-subjective textual or image-based requirements~\cite{regenwetter2025multi, zhong2023topology}. For example, one of the bike company's customers may be able to request a ``futuristic black cyberpunk-style road bike''~\cite{regenwetter2025multi} to be custom designed by the personalization tool. Modern vision-language models (VLMs) are becoming increasingly capable at engineering-based problem-solving~\cite{picard2024conceptmanufacturingevaluatingvisionlanguage,lai2024visionlanguagemodelbasedphysicalreasoning}. Soon, GenAI may prove a reliable and effective method to translate both subjective and quantifiable objectives specified through text, audio, or images into the rigorous mathematical constraints expected by optimization algorithms. If successful, this generative optimization approach would overcome a key shortcoming of optimization methods seen in empirical studies~\cite{saadi2023generative, saadi2024effect}, expanding both the accessibility and utility of computational problem-solving tools.

\subsubsection{Inferring Requirements from Data}
Constraints and objectives can conceivably be inferred from previous solutions, making their definition in optimization even easier, a technique known as inverse optimization~\cite{chan2025inverse}. Inferring some or all constraints using inverse optimization can ameliorate the sometimes tedious~\cite{saadi2023generative} and intractable~\cite{fajemisin2024optimization} process of specifying constrains manually and exactly. This implicit inference is essential in the absence of explicitly-specified constraints, otherwise optimization generally struggles to generate feasible solutions~\cite{regenwetter2025bike}.

Although GenAI is more reliable than optimization at implicitly satisfying constraints~\cite{regenwetter2025bike}, optimization-inspired training methods for implicit learning can further improve constraint satisfaction in engineering problems, including bike design~\cite{regenwetter2024constraining}. Alternatively, the bike manufacturer can improve the robustness of optimization by applying insights from existing data through 
GenAI's distribution modeling principles~\cite{seshadri2016density, cook2018horsetail, chen2024data, dapogny2023entropy, kapteyn2019distributionally}. This technique has also yielded demonstrably better bicycle design constraint satisfaction~\cite{regenwetter2025multi}. This family of generative optimization approaches might allow an autonomous bike optimization system to use implicit knowledge from the distribution of existing designs or customer feedback of those designs, to infer less-salient bicycle design requirements~\cite{regenwetter2022framed} left unspecified by the user.

\subsubsection{Unifying Solution Representations} 
As generative optimization seeks to accommodate the many diverse ways of defining problems, it is equally important to address the many ways of defining solutions. Pure optimization approaches typically require rigidly defined solution representations~\cite{martins2021engineering}, despite the fact that different solutions may have different optimal representations. For example, a bicycle component to be 3D-printed could easily be represented with a surface mesh~\cite{kumar2023printing}, but a machined component would ideally be parametrically defined~\cite{colligan2022hierarchical}.

In stark contrast to optimization, multimodality and flexible, mixed length data representation has enabled some of the most noteworthy advancements in GenAI~\cite{radford2021learning, achiam2023gpt}. Representation learning techniques can learn translations between solution modalities for use in optimization~\cite{jain2023vectorfusion, frans2022clipdraw, michel2022text2mesh}, including engineering optimization~\cite{regenwetter2025multi, rios_10371898, wong2024generative}.

Notably, incorporating multimodal GenAI models into optimization carries key limitations, many echoed in previous sections, such as cost~\cite{samsi2023words, zhou2024survey} or need for retraining~\cite{jiang2019free, tripp2020sampleefficient}. Additionally, direct optimization in embedding spaces often fails to converge, due to model biases~\cite{michel2022text2mesh, frans2022clipdraw}. As GenAI-based multimodality becomes more prominent in engineering problems~\cite{Song2023Multimodal, yu2025gencad}, addressing key limitations will likely enable more powerful multimodal 
generative optimization methods in the near future. Such models could enable the bike company's personalization tool to seamlessly integrate textual or image based initial solutions, bridge gaps to the mesh-based representations needed for numerical simulations, and ultimately output a final design solution in a parametric format for easy manufacturing.

\section{Conclusion}\label{sec13}
Generative optimization is a compelling integration of the robust capabilities of traditional optimization with the speed and flexibility of GenAI.
Optimization can help GenAI models create precise, high-quality solutions that reliably satisfy requirements.
In turn, GenAI models can accelerate optimization, help it learn implicit constraints, or alleviate its dependence on rigidly defined problem or solution spaces. Even in its infancy, generative optimization is beginning to overcome key limitations of existing computational problem-solving formulations, tackling previously-intractable problems with real-time speed and flexible control.  

Despite early successes, many generative optimization approaches have countless challenges to overcome with precision, reliability, computational cost, and more, in order to realize their significant untapped potential.
With ongoing developments and research, generative optimization stands on the cusp of pioneering innovations in engineering. The future for this technology looks promising, with the potential to drive significant breakthroughs in how complex problems are approached and solved.

\backmatter

\bibliography{biblio}%

\end{document}